\documentclass[prl,nofootinbib]{revtex4}

\usepackage{exscale}

\usepackage{amssymb,amsbsy}

\usepackage[dvips]{graphicx}

\newcommand{\bea}{\begin{eqnarray}}

\newcommand{\eea}{\end{eqnarray}}

\newcommand{\pn}{$\pi^0\pi^0$ }

\newcommand{\pc}{$\pi^{\pm}\pi^0$ }

\newcommand{\bc}{\begin{center}}

\newcommand{\ec}{\end{center}}

\newcommand{\bfi}{\begin{figure}}

\newcommand{\efi}{\end{figure}}

\newcommand{\igr}{\includegraphics}

\def\bpi{\boldsymbol\pi}

\begin{document}

\title{Low-energy $\bpi\bpi$ photoproduction off nuclei}

\author{P. M\"uhlich}


\author{L. Alvarez-Ruso}

\author{O. Buss}

\author{U. Mosel}

\affiliation{Institut f\"ur Theoretische Physik, Universit\"at Giessen, D--35392 Giessen, Germany}

\begin{abstract}

In the present paper we investigate \pn and \pc photoproduction off complex nuclei at incident beam energies of $400-460$ MeV. Simulations of two pion photoproduction on protons and nuclei are performed by means of a semi-classical BUU transport model including a full coupled-channel treatment of the final state interactions. Elastic scattering of the final state pions with the nucleons in the surrounding nuclear medium is found to yield a downward shift of the $\pi\pi$ invariant mass distribution. We show that the target mass dependence of the \pn invariant mass spectrum as measured by the TAPS collaboration can be explained without introducing medium effects beyond absorption and quasi-elastic scattering of the final state particles. On the other hand, we find considerable discrepancies with the data in the \pc channel, which are not understood. 

\end{abstract}

\maketitle

One major goal of current nuclear physics is the observation of at least partial restoration of chiral symmetry. Since the chiral order parameter $\langle\bar qq\rangle$ is expected to decrease by about 30\% already at normal nuclear matter density \cite{Drukarev:1990kd,Cohen:1992nk,Li:1994mq,Brockmann:1996iv}, any in-medium change due to the dropping quark condensate should in principle be observable in photonuclear reactions. The conjecture that such a partial restoration of chiral symmetry causes a softening and narrowing of the $\sigma$ meson as the chiral partner of the pion in the nuclear medium \cite{Bernard:1987im,Hatsuda:1999kd} has led to the idea of measuring the \pn invariant mass distribution near the $2\pi$ threshold in photon induced reactions on nuclei \cite{Schadmand:2002tp}. In contrast to its questionable nature as a proper quasiparticle in vacuum, the $\sigma$ meson might develop a much narrower peak at finite baryon density due to phase-space  suppression for the $\sigma\rightarrow\pi\pi$ decay, hence making it possible to explore its properties when embedded in a nuclear many-body system \cite{Schuck:1988jn,Aouissat:1995sx,Chiang:1998di,Jido:2000bw}. Measuring a threshold enhancement of the \pn invariant mass spectrum might serve as a signal for the partial restoration of chiral symmetry inside nuclei and, therefore, give information about one of the most fundamental features of QCD.

A first measurement of the two pion invariant mass spectrum has been obtained by the CHAOS collaboration in pion induced reactions on nuclei \cite{Bonutti:1996ij,Bonutti:2000bv}. An accumulation of spectral strength near the $2\pi$ threshold has been observed for large nuclei in the $\pi^+\pi^-$ channel. This can be interpreted as a change of the $\pi\pi$ interaction or, respectively, of the properties of the $\sigma$ meson in the nuclear environment. The photon can test much higher densities than any hadronic probe due to its electromagnetic coupling to the nucleon  and, therefore, is much better suited to investigate the in-medium properties of hadrons. A recent experiment with the TAPS spectrometer at the tagged-photon facility MAMI-B in Mainz shows an even more pronounced accumulation of spectral strength of the two pion mass spectrum for low invariant masses with increasing target mass corresponding to increasing average densities \cite{Messchendorp:2002au}. This accumulation has been observed for the \pn but not for the \pc final state, which could be interpreted as a significant modification of the $\pi\pi$ interaction in the $\sigma(I=0)$ channel.

One of the present great challenges of nuclear theory is to be able to disentangle new phenomena like in-medium modifications from more conventional nuclear effects such as Fermi motion, absorption and rescattering of the final state particles. In this respect, connecting the outcome of theoretical models concerning the intrinsic properties of the QCD excitation spectrum to experimental observables is an important and frequently undervalued issue. In the present paper we study the production and propagation of pion pairs in nuclei in a semi-classical approximation using vacuum matrix elements but taking into account binding energies, Fermi motion and Pauli blocking of the produced final state. If this approach consistently fails to reproduce the experimental measurements, one confidently might introduce new ideas going beyond what has been established so far. In this respect, the present study can be understood as a cross-check regarding the observation of a modification of the $\sigma$ meson in two pion photoproduction off nuclei.

The authors of Ref.~\cite{Roca:2002vd,Oset:2001iy} studied double pion photoproduction off nuclei in a many-body approach, achieving quite impressive agreement with the experimental data. Their results imply a clear modification of the $\pi\pi$ interaction in the medium that can be related to an in-medium change of the properties of the $\sigma$ meson. Within the model of Ref. \cite{Roca:2002vd} the $\sigma$ meson is generated dynamically through the underlying chiral dynamics of the $\pi\pi$ interaction, which is treated in a chiral-unitary approach. However, the final state interactions (FSI) of the pions with the nucleons in the surrounding nuclear medium have been treated as purely absorptive by means of a Glauber-type damping factor calculated along straight line trajectories. In fact, rescattering of the final pions affect the charge of the detected final state (by charge exchange processes) and changes the trajectories. Therefore, FSI could considerably distort the observed invariant mass spectra.

In the present study the FSI are described within a coupled-channel transport model based on the semi-classical Boltzmann-Uehling-Uhlenbeck (BUU) equation. For details of this model we refer the reader to Refs. \cite{Effenberger:1997rc,Teis:1997kx,Muhlich:2002tu} and references therein. However, we divert from the implementation of BUU used in these papers in the fact that baryonic resonances are not explicitly propagated.  
In addition, low energy pion absorption has been implemented by means of the absorptive part of the pion optical potential developed in Refs.~\cite{Nieves:1993ye,Garcia-Recio:1991xa}. The optical potential of Ref.~\cite{Nieves:1993ye} is an extrapolation for low energy pions of the one obtained previously~\cite{Nieves:1993ev} using microscopic many-body techniques. At pion kinetic energies above $T_{\pi} = 85$~MeV, the potential of Ref.~\cite{Garcia-Recio:1991xa} is used. The later is meant to describe the region dominated by the $\Delta$ resonance and matches well with the low energy one. This is similar to what was done in \cite{Roca:2002vd}. The important difference of our approach to a purely absorptive treatment of the FSI is that in an interaction with a nucleon of the target nucleus the primarily produced hadron might not only be absorbed but also might change its momentum in an elastic collision. Furthermore, for instance, a \pn pair need not to be created in the photon-nucleon interaction but also might be produced from a primary $\pi^+\pi^0$ pair followed by a charge exchange reaction during the FSI. One should bear in mind that the assumption of pions moving along classical trajectories might be somewhat crude at the rather low energies we are dealing with here. On the other side, a good description of 
pion-nucleus reactions within cascade models has been achieved for pions of
energies down to 85~MeV~\cite{Engel:1994jh,Salcedo:1988md} and, more recently, of absorption cross sections at energies as low as 30~MeV~\cite{Oliver}.
Ultimately, it would be desirable to incorporate quantum corrections in
semiclassical approaches~\cite{Salcedo:1988tr}.

For the elementary two pion photoproduction process on the nucleon we exploit the model of Refs. \cite{Nacher:2000eq,GomezTejedor:1996pe}.  This model provides us with a reasonable input for the elementary double-pion production processes on nucleons, not only for the total cross sections but also for the differential distributions. This is illustrated in Fig.~\ref{figure01} where the invariant mass distribution for \pn and \pc photoabsorption is compared to the experimental data measured by the TAPS collaboration \cite{Messchendorp:2002au}. 

\bfi[h!]

\bc

\igr[scale=1.3]{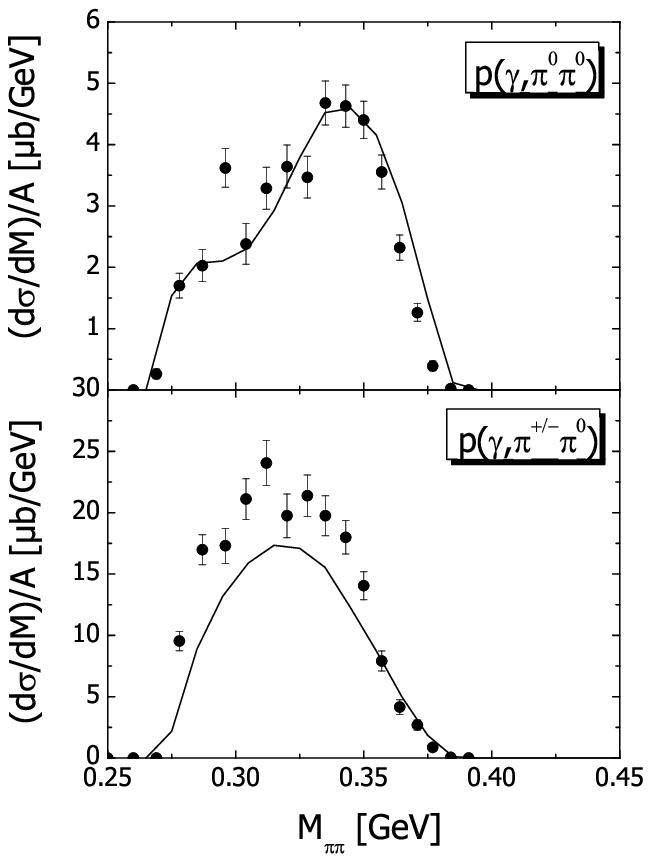}

\caption{Two pion invariant mass distributions for the reactions $p(\gamma,\pi^0\pi^0)$ and $p(\gamma,\pi^{\pm}\pi^0)$ calculated with the model of Ref. \cite{Nacher:2000eq}. The experimental data stem from Ref. \cite{Messchendorp:2002au}.}

\label{figure01}

\ec

\efi

In the following we adopt the coupled-channel transport model in a Monte-Carlo simulation to examine the influence of the FSI on two pion photoproduction off nuclei. To this aim we separate the FSI in absorption, elastic and charge exchange collisions of the produced pions with the nucleons of the target nucleus. In the present calculation we have not included the real part of the pion potential assuming therefore that pions move along straight lines between collisions. So far we have performed preliminary calculations including the Coulomb potential that gives small corrections to observables involving charged pions in the final state. A more complete study including the real part of the pion optical potential will be available in a forthcoming paper~\cite{FutureWork}.

The probability of a pion-nucleon elastic (charge-exchange) collision is determined using the experimental values of the vacuum cross sections for the available channels and assuming isospin symmetry for the rest. If an elastic process occurs, then the momenta of the final particles are sorted according to the vacuum pion-nucleon angular distributions. We get these differential cross sections for all elastic and charge exchange reactions in terms of the phase shifts and inelasticities obtained in the SAID analysis~\cite{Arndt:1995bj,Arndt:2003fj}.Typically, BUU simulations assume isotropic angular distributions in the pion-nucleon center-of-mass frame . As a matter of fact, the results presented here with realistic angular distributions differ only slightly from those obtained with the isotropic prescription. This is in line with the results obtained earlier in Ref.~\cite{Engel:1994jh}.

In Fig. \ref{figure02} we show the two pion invariant mass distributions for the $^{12}C(\gamma,\pi^0\pi^0)$ and the $^{208}Pb(\gamma,\pi^0\pi^0)$ reactions. For the solid curves we assumed purely absorptive FSI, implying that the final state pions do not undergo elastic or charge exchange collisions. These curves can be compared to the dotted lines that represent the theoretical results of Ref. \cite{Roca:2002vd} and \cite{Messchendorp:2002au}~\footnote{The results of the calculation for the \pc channel given in Ref.~\cite{Messchendorp:2002au} should be multiplied by a factor 2~\cite{Rocapriv}.} involving the $\pi\pi$ interaction in the vacuum. Within the theoretical uncertainties of both models these results can be regarded as identical in the \pn channel. However, in the \pc channel  we obtain larger differential cross sections for both nuclei, specially at higher invariant masses. Since we are using the same input for the elementary reaction and the same absorption mechanism as in \cite{Roca:2002vd} we would have expected a similar level of agreement as in the \pn channel. 

\bfi[h!]

\bc

\igr[scale=1.5]{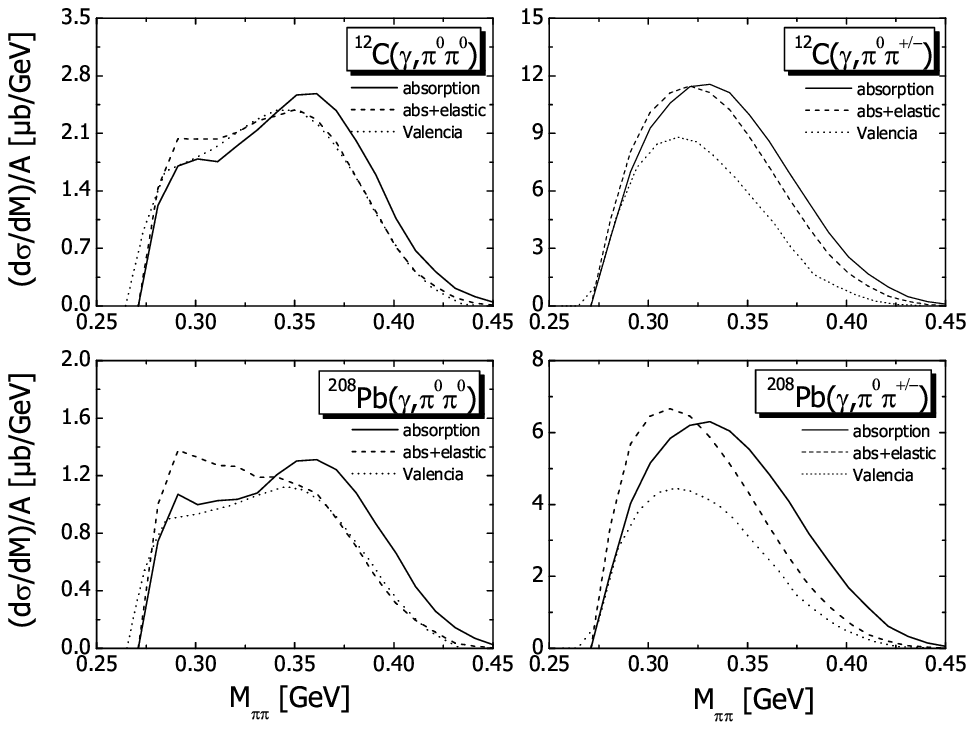}

\caption{Two pion invariant mass distribution for the $A(\gamma,\pi^0\pi^0)$ and $A(\gamma,\pi^{\pm}\pi^0)$ reaction for $A=^{12}$C and $A=^{208}$Pb. For the solid line the FSI consist merely of pion absorption whereas in the calculation yielding the dashed line, the pions are allowed to scatter elastically, i.e. without charge exchange, off the nucleons in the surrounding nuclear medium. The dotted lines show the theoretical results of Ref. \cite{Roca:2002vd,Messchendorp:2002au,Rocapriv} without the medium modification of the $\pi\pi$ final state interaction.}

\label{figure02}

\ec

\efi

Extending the FSI to absorption and elastic collisions without charge exchange we obtain the results depicted by the dashed lines. An appreciable shift of strength to the low invariant mass region is observed. This shift can be attributed to the energy loss of the pions that undergo elastic collisions, in most cases yielding a downward shift of the invariant mass of the respective pion pair. This can at least partially be understood as follows. In general, the two pions of the primary produced $\pi\pi$ pair have unequal momenta. Within the relevant kinematics the mean value of the single pion momenta ranges somewhere between 100 MeV and 150 MeV. In this energy domain the elastic cross section for $\pi N$ scattering varies dramatically, leading to a high probability for the pion with the higher momentum, say above 150 MeV, to scatter elastically. The invariant mass of the pair is a function of the absolute value of the two pion momenta and the angle between the pion moving directions. Keeping the momentum of the slow moving pion and the angle constant, this function rises monotonously with the value of the larger pion momentum. In an elastic $\pi N$ collision the fast pion in average will lose some of its kinetic energy, hence leading to a downward shift of the invariant mass of the pion pair. The change of the angle between the pion directions does not have a net effect, since this angle is changed more or less isotropically, hence leading in some cases to higher and in other cases to lower invariant masses.

In Fig. \ref{figure03} we show the invariant mass distributions including also charge exchange $\pi N$ collisions. The total cross section is enhanced by these reactions. In the \pn channel, this enhancement is due to the fact that the cross section for photoproduction of \pc pairs is much larger than the cross section for \pn photoproduction, hence leading to more side-feeding by the $\pi^+ n\to \pi^0 p$ and $\pi^- p\to\pi^0 n$ reactions as compared to the loss of neutral pion flux by the inverse reactions. An equivalent much more pronounced effect is observed in the \pc channel because the cross section for $\pi^+ \pi^-$ photoproduction is much larger than the cross sections of any of the other channels, hence leading to considerable side-feeding by means of the charge exchange reactions mentioned above.

\bfi[h!]

\bc

\igr[scale=1.5]{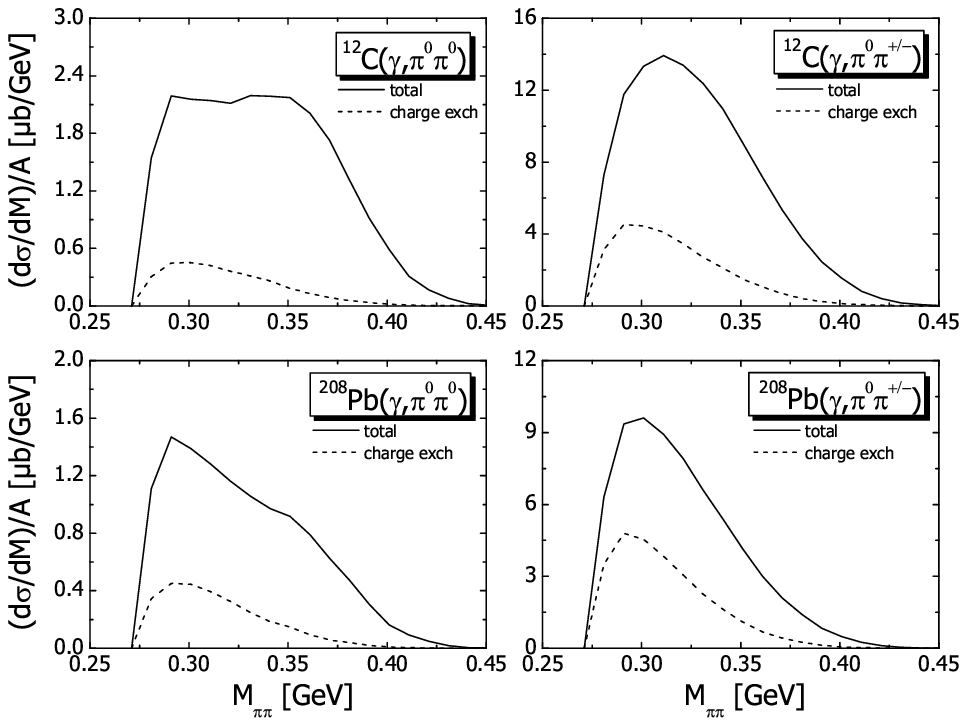}

\caption{Two pion invariant mass distribution for \pn and \pc photoproduction of $A=^{12}$C and $A=^{208}$Pb including $\pi N$ charge exchange collisions. The contribution from the charge-exchange reactions is indicated by the dashed lines.}

\label{figure03}

\ec

\efi  

For the charge exchange collisions, the same kinematic considerations as in the purely elastic case hold, therefore leading to an even more pronounced shift of strength to the low invariant mass region as can be seen from the dashed line in figure \ref{figure03}. We have also calculated the contribution from single pion photoproduction via inelastic $\pi N\to\pi\pi N$ collisions. Within the considered energy range, those pion pairs give an additional correction of about 5\% to the total cross sections and, therefore, have been omitted in the present calculations.

In Fig. \ref{figure04} we present our results for both the $A(\gamma,\pi^0\pi^0)$ and the $A(\gamma,\pi^{\pm}\pi^0)$ reaction in comparison to the experimental data from the TAPS collaboration \cite{Messchendorp:2002au}. In the \pn channel we find excellent agreement with the experimental data. Our curves also look very similar to the results of Ref. \cite{Roca:2002vd} despite the completely different physics involved. The downward shift in the invariant mass spectrum, which in Ref.~\cite{Messchendorp:2002au} was taken as an indication for chiral symmetry restoration or in-medium modification of the $\pi \pi$ correlation, is reproduced solely by ordinary final state interactions of the two outgoing pions.

In the \pc channel our results are considerably higher than the data and also above those obtained by Roca et al.~\cite{Roca:2002vd}. The difference of the two theoretical curves can -- at least partially -- be understood by the incomplete set of FSI incorporated in the model of Ref. \cite{Roca:2002vd}. Side-feeding has been omitted in Ref. \cite{Roca:2002vd} by restricting the interactions of the produced pions with the nuclear medium to purely absorptive FSI. The discrepancy with the experimental data is a standing problem  because the apparently stronger pion absorption in the \pc channel than in the \pn channel as indicated by the experimental measurements (compare Figs. \ref{figure01} and \ref{figure04}) is not understood. Concerning the shapes, our results in the \pc channel are also shifted towards low invariant masses. However, the quality of the available data does not allow to discriminate between the shape we have obtained and one which is not downward shifted, like the one of Ref.~\cite{Roca:2002vd}.

\bfi[h!]

\bc

\igr[scale=1.5]{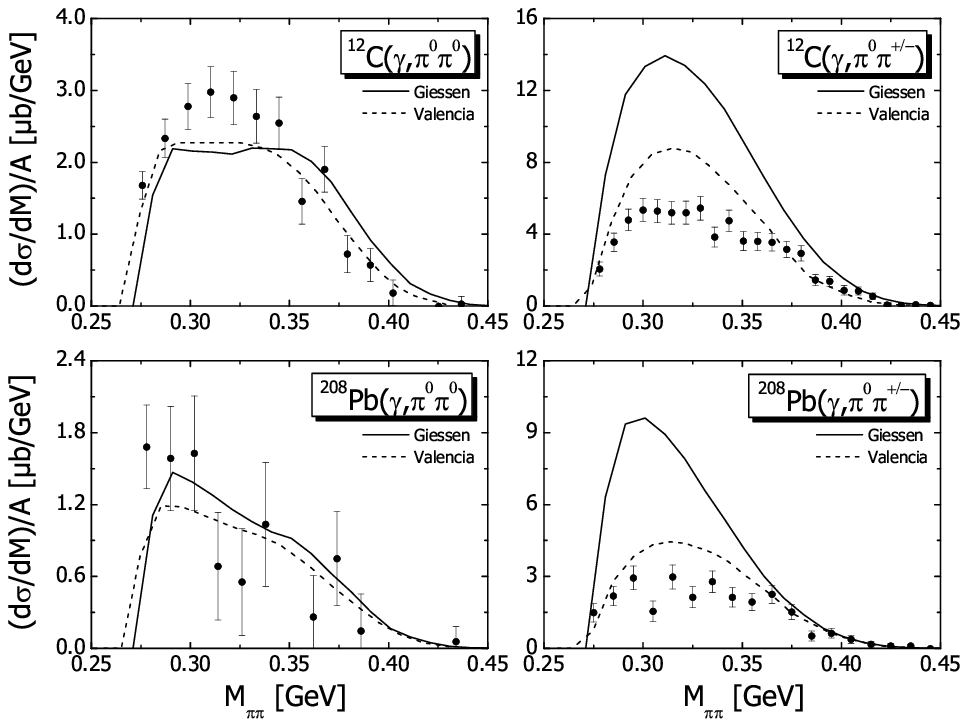}

\caption{Two pion invariant mass distributions for \pn and \pc photoproduction of $^{12}$C and $^{208}$Pb. The continuous lines labeled "Giessen" represent the results of the present paper and the dashed lines labeled "Valencia" depict the results of Ref. \cite{Roca:2002vd} using the in-medium $\pi\pi$ interaction in the \pn channel and of Ref.~\cite{Messchendorp:2002au,Rocapriv} in the \pc channel. The experimental data stem from Ref. \cite{Messchendorp:2002au}.}

\label{figure04}

\ec

\efi

In summary, we have indicated the relevant role of some conventional FSI effects in two-pion photoproduction in nuclei. Elastic scattering of the pions with nucleons of the target nucleus yields a loss of kinetic energy of the scattered pions and therefore, in average, to a reduced invariant mass of the pion pair. Charge exchange reactions produce a non-negligible contribution to the $\pi\pi$ mass spectra, giving rise to an even more pronounced accumulation of spectral strength in the low invariant mass region. On the other hand, side-feeding especially from the $\gamma N\to\pi^+\pi^-N$ reaction followed by a $\pi N$ collision with charge exchange, that cannot be accounted for in models involving purely absorptive FSI, further increases the differential cross sections in the \pc channel, to values significantly above the data. The apparently stronger absorption of pions in the \pc channel as compared to the \pn channel is still an unsolved problem both from the experimental and theoretical point of view. However, our explanation of the observed downward shift of the invariant mass spectrum follows completely well established nuclear physics effects, that, even if a complete understanding of $\pi\pi$ photoproduction off complex nuclei is not yet possible, have to be accounted for in any serious calculation. We therefore conclude that, given our present understanding, the observed target mass dependence of the \pn invariant mass spectrum can be explained by final state interactions and thus does not provide an unmistakable evidence on the modification of the $\pi\pi$ interaction or, respectively, on the properties of the $\sigma$ meson at finite nuclear densities. More work, both experimental and theoretical, is required to clarify this issue.

The authors acknowledge valuable discussions with J. Lehr, V. Metag, M. Post and S. Schadmand. This work was supported by DFG and BMBF. One of us, LAR, has been supported by the Alexander von Humboldt Foundation. He also would like to thank E. Oset, M. J. Vicente Vacas and specially L. Roca for extensive discussions on this problem.

\bibliography{bibliotwopi}

\end{document}